\documentclass[prd,nofootinbib,showpacs,preprint]{revtex4}
\usepackage{amsmath}
\usepackage{graphicx}
\graphicspath{{Figs/}}
\usepackage{dcolumn}
\usepackage{bm}
\usepackage{amssymb}
\usepackage[usenames,dvipsnames]{color}
\usepackage{slashed}
\usepackage[dvipdfm,colorlinks,citecolor=blue]{hyperref}
\begin{document}
\title{Neutrino masses and mixings with non-zero $\theta_{13}$ in \\Type I+II Seesaw Models}
\author{Debasish Borah}
\email{dborah@tezu.ernet.in}
\affiliation{Department of Physics, Tezpur University, Tezpur-784028, India}
\author{Mrinal Kumar Das}
\email{mkdas@tezu.ernet.in}
\affiliation{Department of Physics, Tezpur University, Tezpur-784028, India}


\begin{abstract}
We study the survivability of neutrino mass models with normal as well as inverted hierarchical mass patterns in the presence of both type I and type II seesaw contributions to neutrino mass within the framework of generic left-right symmetric models. At leading order, the Dirac neutrino mass matrix is assumed to be diagonal with either charged lepton (CL) type or up quark (UQ) type structure which gets corrected by non-leading effects giving rise to deviations from tri-bi-maximal (TBM) mixing and hence non-zero value of $\theta_{13}$. Using the standard form of neutrino mass matrix which incorporates such non-leading effects, we parametrize the neutrino mass matrix incorporating both oscillation as well as cosmology data. Also considering extremal values of Majorana CP phases such that the neutrino mass eigenvalues have the structure $(m_1, -m_2, m_3)$ and $(m_1, m_2, m_3)$, we then calculate the predictions for neutrino parameters in the presence of both type I and type II seesaw contributions, taking one of them dominant and the other sub-dominant. We show that these mass models can survive in our framework with certain exceptions.
\end{abstract}
\maketitle
\section{Introduction}
Recent neutrino oscillation experiments have provided significant amount of evidence which confirms the existence of the non-zero yet tiny neutrino masses \cite{PDG}. We know that the smallness of three Standard Model
neutrino masses can be naturally explained 
via seesaw mechanism. In general, such seesaw mechanism can be of three types : type I \cite{ti}, type II \cite{tii} and type III \cite{tiii}. All these mechanisms involve the inclusion of additional fermionic or scalar fields to generate tiny neutrino masses at tree level. Although these seesaw models can naturally explain the smallness of neutrino mass compared to the electroweak scale, we are still far away from understanding the origin of neutrino mass hierarchies as suggested by experiments. Recent neutrino oscillation experiments T2K \cite{T2K}, Double ChooZ \cite{chooz}, Daya-Bay \cite{daya} and RENO \cite{reno} have not only made the earlier predictions for neutrino parameters more precise, but also predicted non-zero value of the reactor mixing angle $\theta_{13}$. The latest global fit value for $3\sigma$ range of neutrino oscillation parameters \cite{schwetz12} are as follows:
$$ \Delta m_{21}^2=(7.00-8.09) \times 10^{-5} \; \text{eV}^2$$
$$ \Delta m_{31}^2 \;(\text{NH}) =(2.27-2.69)\times 10^{-3} \; \text{eV}^2 $$
$$ \Delta m_{23}^2 \;(\text{IH}) =(2.24-2.65)\times 10^{-3} \; \text{eV}^2 $$
$$ \text{sin}^2\theta_{12}=0.27-0.34 $$
$$ \text{sin}^2\theta_{23}=0.34-0.67 $$ 
\begin{equation}
\text{sin}^2\theta_{13}=0.016-0.030
\end{equation}
where NH and IH refers to normal and inverted hierarchy respectively. The best fit value of $\delta_{CP}$ turns out to be $300$ degrees \cite{schwetz12}.

The above recent data have positive evidence for non-zero $\theta_{13}$ as well, which was earlier thought to be zero or negligibly small. Non-zero $\theta_{13}$ can be explained by incorporating various corrections to the standard TBM mixing. The standard TBM mixing pattern is
\begin{equation}
U_{TBM}==\left(\begin{array}{ccc}\sqrt{\frac{2}{3}}&\frac{1}{\sqrt{3}}&0\\
 -\frac{1}{\sqrt{6}}&\frac{1}{\sqrt{3}}&-\frac{1}{\sqrt{2}}\\
-\frac{1}{\sqrt{6}}&\frac{1}{\sqrt{3}}& \frac{1}{\sqrt{2}}\end{array}\right),
\end{equation}
which predicts $\text{sin}^2\theta_{12}=\frac{1}{3}$, $\text{sin}^2\theta_{23}=\frac{1}{2}$ and $\text{sin}^2\theta_{13}=0$. However, since the latest data have ruled out $\text{sin}^2\theta_{13}=0$, there arises the need to go beyond the TBM framework. Since the experimental value of $\theta_{13}$ is still much smaller than the other two mixing angles, TBM can still be a valid approximation and the non-zero $\theta_{13}$ can be accounted for by incorporating non-leading contributions to TBM coming from charged lepton mass diagonalization, for example. There have already been a great deal of activities in this context \cite{nzt13, nzt13GA} and the latest data can be successfully predicted within the framework of several interesting models. These frameworks which predict non-zero $\theta_{13}$ may also shed light on the Dirac CP violating phase which is still unknown (and could have remained unknown if $\theta_{13}$ were exactly zero).

Apart from predicting the correct neutrino oscillation data as well as the Dirac CP phase, the nature of neutrino mass hierarchy is also an important yet unresolved issue. Use of specific grand unified models explaining the seesaw mechanisms has also been done in the last few years to study the hierarchy of neutrino masses. An analysis done in \cite{CHA} showed that every normal neutrino mass hierarchy solution of a grand unified model corresponds to an inverted hierarchy solution. It was also mentioned in their work that any future observation of inverted hierarchy would tend to disfavor the grand unified models based on the conventional type I seesaw mechanism. But models with type II and type III or models based on conserved $ L_e-L_{\mu}-L_{\nu}$ symmetry may favor the inverted hierarchical nature of neutrino masses. Models based on seesaw mechanism with three right handed neutrinos can also generate inverted hierarchical neutrino masses \cite{NNS} within the framework of bi-maximal mixing. Understanding the correct nature of hierarchy can also have non-trivial relevance in leptogenesis as well as cosmology. For example, the latest cosmology data on the sum of absolute neutrino masses \cite{SFO} have already ruled out the scenario of quasi-degenerate (QDN) neutrino masses with $m_i \geq 0.1 \; \text{eV}$. From supernova neutrinos point of view, it was shown \cite{HM} that one can discriminate the inverted hierarchy from the normal one if $\text{sin}^2\theta_{13}\geq \text{a few} \times 10^{-4}$. If a particular neutrino mass hierarchy is assumed this can bias cosmological parameter constraints \cite{PRD80} like dark energy equation of state parameter as well as the sum of the neutrino masses. Therefore, the study of normal and inverted hierarchy using different types of seesaw mechanism is very important both from neutrino physics and cosmology point of view.

In view of above, the present work is planned to carry out a study of neutrino mass models with normal and inverted hierarchical neutrino masses in the framework of generic left-right symmetric models (LRSM) \cite{lrsm}. Such a work was done recently in \cite{mkd-db-rm} where TBM type $\mu-\tau$ symmetric neutrino mass matrix was considered. In that study, the dominating type I seesaw term was kept fixed and type II term was varied and the predictions for neutrino oscillation parameters were calculated. In this present work, we include non-zero $\theta_{13}, \delta_{CP}$ as well as to consider the case where type II seesaw term dominates over type I. We parametrize the neutrino mass matrix for QDN scenario using global fit neutrino oscillation as well as cosmology data. The dominating seesaw term is then used to find out the right-handed Majorana neutrino mass matrix by assuming two different types of Dirac neutrino mass matrices (CL and UQ type) and two different Majorana CP phase patterns ($(m_1,-m_2,m_3)$ and $(m_1,m_2,m_3)$). Fixing the dominating seesaw term in this way, the other seesaw term is allowed to vary and predictions are calculated for all these variations. By calculating the predictions for neutrino oscillation parameters, we show that both type I dominating and type II dominating cases give almost identical results. We find that normal hierarchy along with CL type Dirac neutrino mass matrix gives rise to predictions within $3\sigma$ range of experimental data for both types of Majorana CP phase patterns. Whereas, inverted hierarchy with CL type Dirac mass matrix can exist only with the Majorana CP phase pattern $(m_1,-m_2,m_3)$. Also, we show that UQ type Dirac neutrino mass matrix is disfavored for both the types of hierarchies and Majorana CP phases.

This paper is organized as follows: in section \ref{method} we discuss the methodology of type II seesaw mechanism in generic LRSM. In section \ref{numeric} we discuss our numerical analysis and results. We then finally conclude in section \ref{conclude}.
\section{Type II seesaw in LRSM}
\label{method}
Type I seesaw framework is the  simplest mechanism for generating tiny neutrino masses and mixing. Such a mechanism is possible in extensions of the standard model by three right handed neutrinos. There is also another type of non-canonical seesaw formula (known as type-II seesaw formula)\cite{tii} where  a left-handed Higgs triplet $\Delta_{L}$ picks up a vacuum expectation value (vev). This is possible both in the minimal extension of the standard model by $\Delta_{L}$ or in other well motivated extensions like left-right symmetric models (LRSM) \cite{lrsm}. The seesaw formula in LRSM can be written as
\begin{equation}
m_{LL}=m_{LL}^{II} + m_{LL}^I
\label{type2a}
\end{equation}
 where the usual  type I  seesaw formula  is given by the expression,
\begin{equation}
m_{LL}^I=-m_{LR}M_{RR}^{-1}m_{LR}^{T}.
\end{equation}
Here  $m_{LR}$ is the Dirac neutrino mass matrix. The above seesaw formula with both type I and type II contributions can naturally arise in extension of standard model with three right handed neutrinos and one copy of $\Delta_{L}$. However, we will use this formula in the framework of LRSM where $M_{RR}$ arises naturally as a result of parity breaking at high energy and both the type I and type II terms can be written in terms of $M_{RR}$. In LRSM with Higgs triplets, $M_{RR}$ can be expressed as $M_{RR}=v_{R}f_{R}$ with $v_{R}$ being the vev of the right handed triplet Higgs field $\Delta_R$ imparting Majorana masses to the right-handed neutrinos and $f_{R}$ is the corresponding Yukawa coupling. The first term $m_{LL}^{II}$ in equation (\ref{type2a}) is due to the vev of $SU(2)_{L}$ Higgs triplet. Thus, $m_{LL}^{II}=f_{L}v_{L}$ and $M_{RR}=f_{R}v_{R}$, where $v_{L,R}$ denote the vev's and $f_{L,R}$ are symmetric $3\times3$ matrices. The left-right symmetry demands $f_{R}=f_{L}=f$. The induced vev for the left-handed triplet $v_{L}$ can be shown for generic LRSM to be
$$v_{L}=\gamma \frac{M^{2}_{W}}{v_{R}}$$
with $M_{W}\sim 80.4$ GeV being the weak boson mass such that 
$$ |v_{L}|<<M_{W}<<|v_{R}| $$ 
In general $\gamma$ is a function of various couplings in the scalar potential of generic LRSM and without any fine tuning $\gamma$ is expected to be of the order unity ($\gamma\sim 1$). type-II seesaw formula in equation (\ref{type2a}) can now be expressed as
\begin{equation}
m_{LL}=\gamma (M_{W}/v_{R})^{2}M_{RR}-m_{LR}M^{-1}_{RR}m^{T}_{LR}
\label{type2}
\end{equation}

With above seesaw formula (\ref{type2}), the neutrino mass matrices
are constructed by considering contributions from both type I and type II terms. The choice of $v_R$ however, remains ambiguous in the literature where different choices of $v_{R}$ are made according to convenience \cite{bstn,bd,ca,aw}. However, in this present work we will always take $v_{R}$ as $v_R=\gamma\frac{M^2_W}{v_L} \simeq \gamma \times 10^{15}\;\text{GeV}$ \cite{aw}. It is worth mentioning that, here $SU(2)_R \times U(1)_{B-L}$ gauge symmetry breaking scale (as in generic LRSM) $v_R$ is the same as the scale of parity breaking \cite{bstn}. Using this form of $v_R$, the seesaw formula (\ref{type2}) becomes 
\begin{equation}
m_{LL}=\gamma \left (\frac{M_{W}}{\gamma \times 10^{15}} \right )^{2}M_{RR}-m_{LR}M^{-1}_{RR}m^{T}_{LR}
\label{type2b}
\end{equation}

Quantitatively, either of the two terms on the right hand side of equation (\ref{type2b}) can be dominant or both the terms can be equally dominant. However, for generic choices of symmetry breaking scales (mentioned above) as well as the Dirac neutrino mass matrices (generically to be of same order as corresponding charged lepton masses), both type I and type II term can be equally dominant only when the dimensionless parameter $\gamma$ is fine tuned to be very small. We check this by equating both the terms to the best fit value of $m_{LL}$. We skip such a fine-tuned case here and consider two other possible cases in our work: one in which type I term dominates whereas type II term is present as a small perturbation and the other in which type II term dominates with type I term as a small perturbation.

\subsection{Case I: Dominating Type I Seesaw}
In this case, the second term in the equation (\ref{type2b}) gives the leading contribution to $m_{LL}$ and hence we compute the right-handed neutrino mass matrix $M_{RR}$ by using the inverse type I seesaw formula $M_{RR}=m_{LR}^Tm_{LL}^{-1}m_{LR}$ where we use the best fit $m_{LL}$ and generic $m_{LR}$ as will be shown in the section \ref{numeric}. Here we hold $M_{RR}$ fixed, so the first term in equation (\ref{type2b}) is dependent on the value of $\gamma$ while second term is fixed. For $\gamma \sim 1$, the first term has minimum contribution whereas for smaller values of $\gamma$, the contribution of the type II term will increase. We vary the dimensionless parameter $\gamma$ from $0.001$ to $1.0$  and check the survivability of neutrino mass models with contributions from type I and type II terms. We adopt a \textit{natural selection} for the survival of neutrino mass models which have the least deviation of $\gamma$ from unity. Nearer the value of $\gamma$ to one, better the chance for the survival of the model in question. Thus the value of  $\gamma$ is an important parameter for the proposed natural selection of the neutrino mass models in question.
\begin{table}
\centering
\caption{Global best fit Input parameters and predictions for neutrino mass eigenvalues}
\vspace{0.5cm}
{\small
\begin{tabular}{|c|c|c|c|c|}
 \hline
   Parameters & IH(+-+) &  IH(+++) &  NH(+-+)&  NH(+++)\\ \hline
$\Delta m_{21}^2[10^{-5}\; \text{eV}^2]$&7.50&7.50&7.50&7.50\\  \hline
$\lvert \Delta m_{31}^2\rvert[10^{-3} \; \text{eV}^2]$&2.43&2.43&2.47&2.47\\  \hline
$\text{sin}^2\theta_{12}$&0.340&0.340&0.340&0.340\\  \hline
$\text{sin}^2\theta_{23}$ &0.606 &0.606 &0.606 & 0.606 \\ \hline
$\text{sin}^2\theta_{13}$&0.022&0.022&0.022&0.022\\  \hline
$ \text{sin}^2\delta_{CP}$ &0.75&0.75&0.75&0.75 \\ \hline
$m_3\;(\text{eV})$&0.075&0.075&0.090&0.101\\  \hline
$m_2\; (\text{eV})$&-0.090&0.090&-0.075&0.089\\  \hline
$m_1\;(\text{eV})$&0.089&0.089&0.075&0.088\\  \hline
$\sum_i m_i\;(\text{eV})$&0.25&0.25&0.24&0.28\\  \hline
\end{tabular}
}
\label{table:results1}
\end{table}
\begin{table}[ht]
\centering
\caption{Predictions for neutrino parameters at $\gamma = 1.0$ for the case of Inverted Hierarchy (IH)} 
\vspace{0.5cm}
{\small
\begin{tabular}{|c|c|c|c|c|c|c|c|c|}
 \hline
   Parameters & Case I & Case II & Case I & Case II & Case I & Case II & Case I & Case II \\
              & CL &  CL & CL & CL & UQ & UQ & UQ &  UQ \\ \hline
                            & (+-+) &(+-+) &(+++) & (+++) & (+-+) & (+-+) & (+++) & (+++) \\ \hline

$\Delta m^2_{21}[10^{-5}\;\text{eV}^2]$ & \textcolor{red}{8.11} &8.09& 7.91 & 7.90 & \textcolor{red}{9.01} & \textcolor{red}{8.95} & \textcolor{red}{9.16} & \textcolor{red}{9.08} \\ \hline
$ \Delta m^2_{23} [10^{-3}\;\text{eV}^2] $ & 2.43 & 2.43 & 2.43 & 2.43 & 2.43 & 2.43 & 2.45 & 2.44  \\ \hline   
$\text{sin}^2\theta_{23}$&0.550& 0.550 &0.464 & 0.468&0.549 & 0.549 &0.354 & 0.360 \\  \hline
$\text{sin}^2\theta_{12}$&0.340&0.340&0.328&0.328 & 0.340 & 0.340 & 0.307 & 0.308 \\  \hline
$\text{sin}^2\theta_{13}$&0.021& 0.021 & \textcolor{red}{0.031}& 0.022 & 0.021& 0.021 & \textcolor{red}{0.053} & \textcolor{red}{0.052}  \\  \hline
$ \text{sin}^2\delta_{CP}$ & 0.688 & 0.688 & 0.575 & 0.583 & 0.693 & 0.693 & 0.351 & 0.360 \\ \hline
\end{tabular}
}
\label{table:results2}
\end{table}
\begin{table}[ht]
\centering
\caption{Predictions for neutrino parameters at $\gamma = 1.0$ for the case of Normal Hierarchy (NH)}
\vspace{0.5cm}
{\small
\begin{tabular}{|c|c|c|c|c|c|c|c|c|}
 \hline
   Parameters & Case I & Case II & Case I & Case II & Case I & Case II & Case I & Case II \\
              & CL &  CL & CL  & CL & UQ  & UQ & UQ &  UQ \\ \hline
                            &(+-+) &(+-+) &(+++) & (+++) &(+-+) &(+-+) & (+++) & (+++) \\ \hline
$\Delta m^2_{21}[10^{-5}\;\text{eV}^2]$ & 7.27 &7.28& 7.79 & 7.78 & \textcolor{red}{6.77} & \textcolor{red}{6.81} & \textcolor{red}{8.44} & \textcolor{red}{8.39} \\ \hline
$ \Delta m^2_{31} [10^{-3}\;\text{eV}^2] $ & 2.47 & 2.47 & 2.47 & 2.47 & 2.47 & 2.47 & 2.48 & 2.48  \\ \hline   
$\text{sin}^2\theta_{23}$&0.550& 0.550 &0.546 & 0.546&0.549 & 0.549 &0.539 & 0.540 \\  \hline
$\text{sin}^2\theta_{12}$&0.340&0.340&0.277&0.279 & 0.340 & 0.340 & \textcolor{red}{0.201} & \textcolor{red}{0.205} \\  \hline
$\text{sin}^2\theta_{13}$&0.021& 0.021 & 0.021& 0.021 & 0.021& 0.021 & 0.021 & 0.021  \\  \hline
$ \text{sin}^2\delta_{CP}$ & 0.687 & 0.687 & 0.684 & 0.684 & 0.691 & 0.691 & 0.684 & 0.684 \\ \hline
\end{tabular}
}
\label{table:results3}
\end{table}

\subsection{Case II: Dominating Type II seesaw}
This is the case where the first term in the equation (\ref{type2b}) gives the leading contribution to $m_{LL}$. This scenario within grand unified models like $SO(10)$ have been discussed in \cite{t2GUT}. In this case, we compute $M_{RR}/\gamma$ from the inverse type II seesaw formula
\begin{equation}
\frac{M_{RR}}{\gamma} = \left ( \frac{1\times 10^{15}}{M_W} \right )^2 m_{LL}
\label{t2dm}
\end{equation}
It should be noted that here we are keeping $M_{RR}/\gamma$ constant instead of just $M_{RR}$ as in the case I. Thus, although the first term in equation (\ref{type2b}) remains constant, the second term varies as we vary $\gamma$ and have the minimal contribution for $\gamma \sim 1$. Similar to the case I, here also we vary $\gamma$ between $0.001$ and $1.0$ and calculate the predictions for neutrino parameters. The details of both these case will be presented in details in the next section.

\section{Numerical Analysis and Results}
\label{numeric}

For the purpose of numerical analysis we use the following parametrization of neutrino mass matrix which is a combination of  TBM version of $\mu-\tau$ symmetric matrix with the addition of correction terms coming from charge lepton mass matrix diagonalization \cite{nzt13GA}.
\begin{equation}
m_{LL}=\left(\begin{array}{ccc}
x& y-w&y+w\\
y-w& x+z+w & y-z \\
y+w & y-z& x+z-w
\end{array}\right)
\label{matrix1}
\end{equation}
where $w$ denotes the deviation of $m_{LL}$ from that within TBM frameworks and setting it to zero, the above matrix boils down to the familiar $\mu-\tau$ symmetric matrix. The eigenvalues of this matrix are $m_1 = x+2y, \; m_2 = x-y+z-\sqrt{z^2+3w^2} $ and $m_3 = x-y+z+\sqrt{z^2+3w^2} $

Then we parameterize the above matrix for QDN case. From presently available cosmological constraints, the upper bound on sum of neutrino masses has come down to the lowest value $\sum_im_i\le0.28\;\text{eV}$ \cite{SFO} which has ruled out QDN neutrino models with $m_i\ge 0.1 \;\text{eV}$. Parametrization  of the matrix (\ref{matrix1}) is done with this upper bound and taking the largest allowed value $m_i\le 0.1 \;\text{eV}$ consistent with the latest cosmological data. A classification for three-fold QDN neutrino masses \cite{a} with maximum Majorana CP violating phase in their eigenvalues is used here. CP phase patterns in the mass eigenvalues for both NH and IH are taken as: $(m_1, -m_2, m_3)$ (denoted as $(+-+)$) and $(m_1, m_2, m_3)$ (denoted as $(+++)$). Using the best fit values of the global neutrino oscillation observational data \cite{schwetz12} on solar and atmospheric neutrino mass squared differences, and taking $m_i\le 0.1 \; \text{eV}$, predictions for parameters in the matrix (\ref{matrix1}) are calculated as shown in table \ref{table:results1}. It should be noted that we are denoting our numerical estimates as predictions, which are valid only under the specific assumptions we are taking about the structure of Dirac neutrino mass matrices, the scale of left-right symmetry breaking and the structure of $m_{LL}$ as mentioned. After making this set of choices, we are left with the freedom of choosing the dimensionless parameter $\gamma$ in equation (\ref{type2}). Here we show our results with respect to the variation of this parameter.

After parameterizing the neutrino mass matrix using oscillation and cosmology data, we consider the two cases I and II mentioned in the previous section one by one. First, we consider the case I i.e., type I dominance and calculate the right-handed Majorana neutrino matrix $M_{RR}$ using the inverse type I seesaw formula 
\begin{equation}
M_{RR}=m_{LR}^Tm_{LL}^{-1}m_{LR}, 
\end{equation}
To calculate the $M_{RR}$ for each case, we need to have the Dirac neutrino mass matrix $(m_{LR})$. We take the Dirac neutrino mass matrix $m_{LR}$ to be diagonal at leading order (LO) with either charged lepton mass structure up quark mass structure. The general form of Dirac neutrino mass at LO is  
\begin{equation}
m^{(0)}_{LR}=\left(\begin{array}{ccc}
\lambda^m & 0 & 0\\
0 & \lambda^n & 0 \\
0 & 0 & 1
\end{array}\right)m_f
\label{mLR1}
\end{equation}
where $m_f$ corresponds to $m_\tau \tan{\beta}$ for $(m, n) = (6, 2), \; \tan{\beta} = 40$ in case of charged lepton and $m_t$ for $(m, n) = (8, 4)$ in the case of up-quarks \cite{dm,mkd}. $\lambda = 0.22$ is the standard Wolfenstein parameter. Just like the charged lepton mass matrix (which is diagonal at LO) gets corrected by non-leading terms, we also consider similar non-leading contributions to the Dirac neutrino mass matrix. Such non-leading contribution can be written as
\begin{equation}
\delta m^{(1)}_{LR}=\left(\begin{array}{ccc}
\mathcal{O}(\lambda^m) & \mathcal{O}(\lambda^m) & \mathcal{O}(\lambda^m)\\
\mathcal{O}(\lambda^n) & \mathcal{O}(\lambda^n) & \mathcal{O}(\lambda^n) \\
\mathcal{O}(1) & \mathcal{O}(1) & \mathcal{O}(1)
\end{array}\right)m_f \xi 
\label{mLR2}
\end{equation}
where $\xi$ parametrizes the departure from LO approximation. Since the parameter $w$ in the neutrino mass matrix (\ref{matrix1}) parametrizes the deviation from LO approximation (which is TBM for generic $A_4$ flavor symmetric models), we assume $\xi \sim w$ for our calculation. Now, using $m_{LR} = m^{(0)}_{LR}+\delta m^{(1)}_{LR} $ in the inverse type I seesaw formula above, we calculate $M_{RR}$. Using the values of $M_{RR}$ in the equation (\ref{type2}), we check the variations of mass squared differences as well as the mixing angles with respect to $\gamma$ (varying from $0.001$ to $1$). 
\begin{figure}[ht]
 \centering
\includegraphics{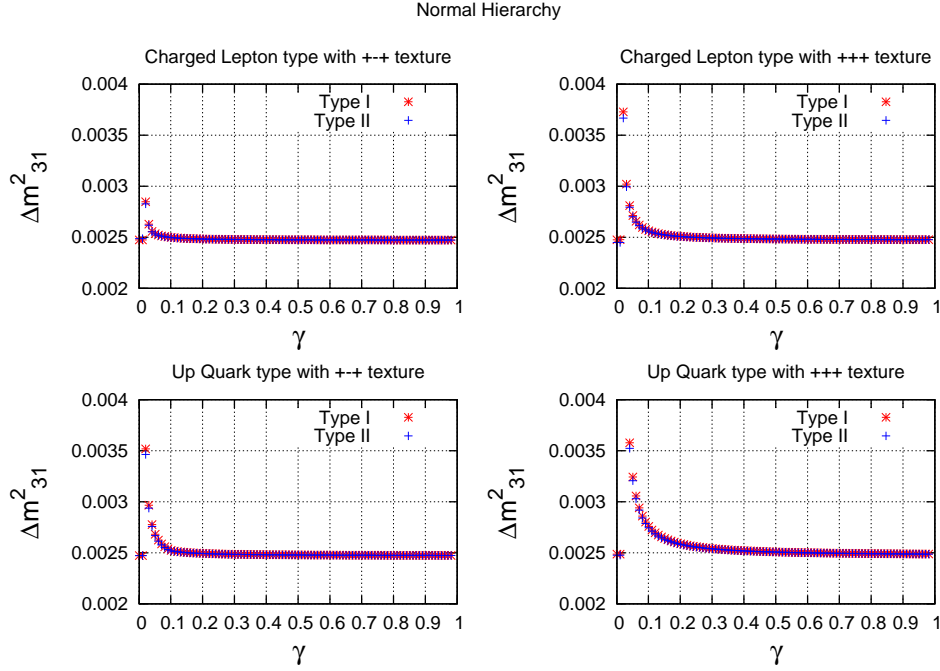}
\caption{Variation of the predicted values of $\Delta m^2_{31}$ as a function of $\gamma$ in NH case}
\label{fig1}
\end{figure}
\begin{figure}[ht]
 \centering
\includegraphics{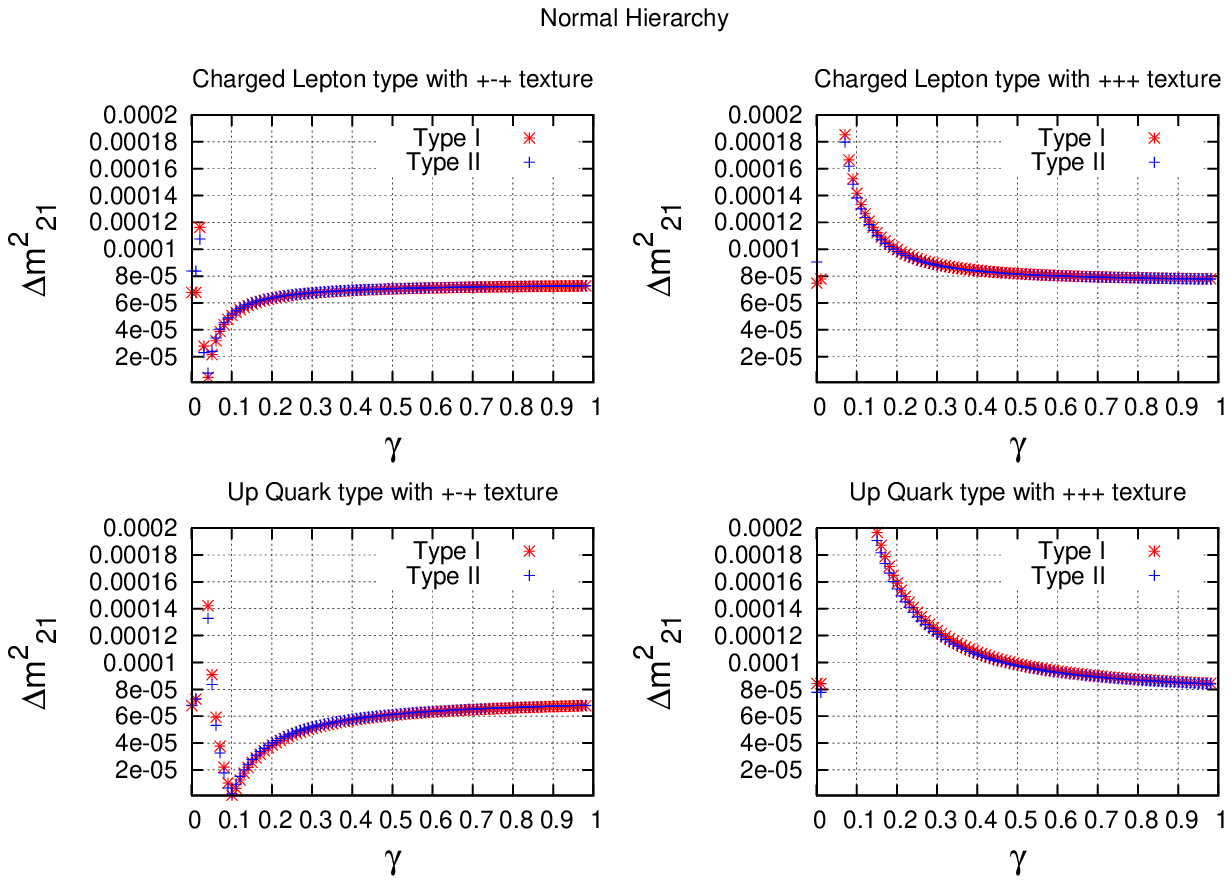}
\caption{Variation of the predicted values of $\Delta m^2_{21}$ as a function of $\gamma$ in NH case}
\label{fig2}
\end{figure}
\begin{figure}[ht]
 \centering
\includegraphics{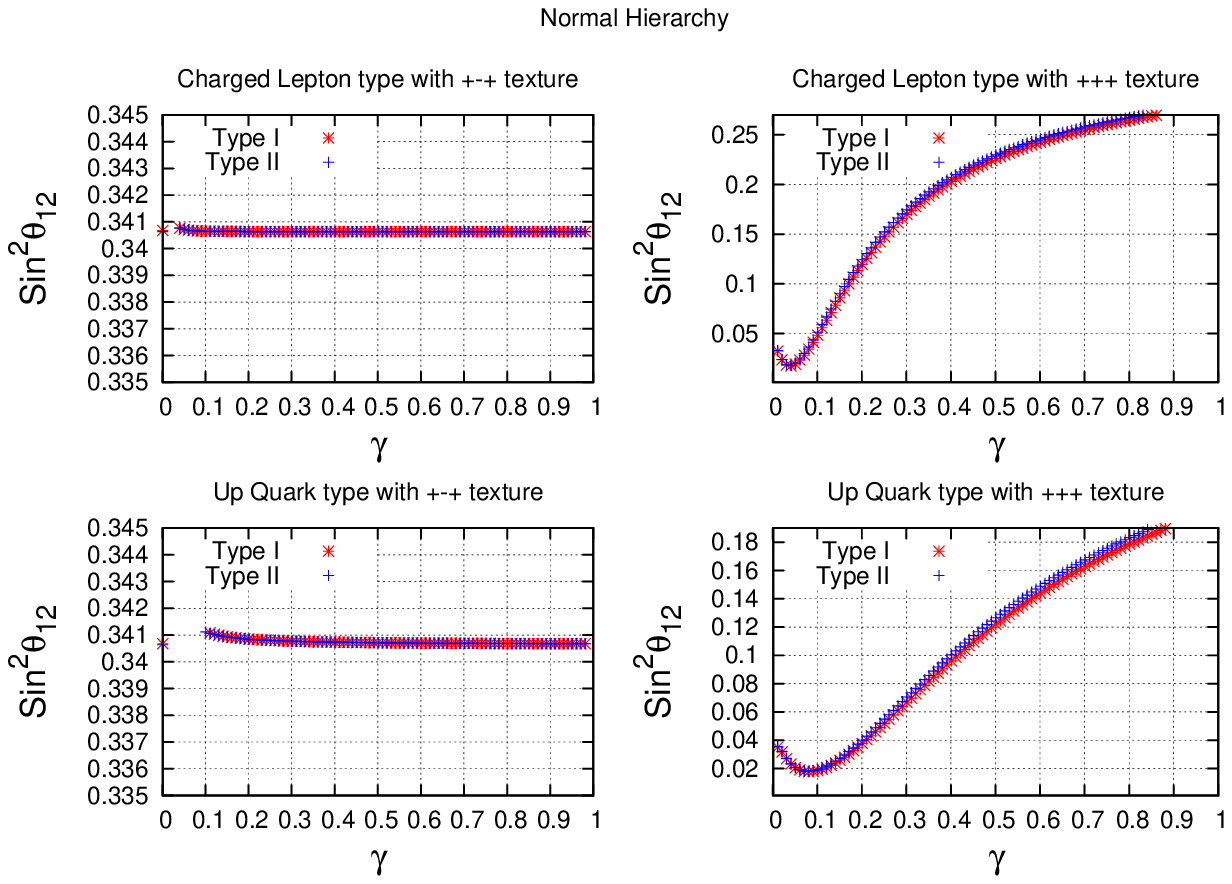}
\caption{Variation of the predicted values of $\sin^2{\theta_{12}}$ as a function of $\gamma$ in NH case}
\label{fig3}
\end{figure}
\begin{figure}[ht]
 \centering
\includegraphics{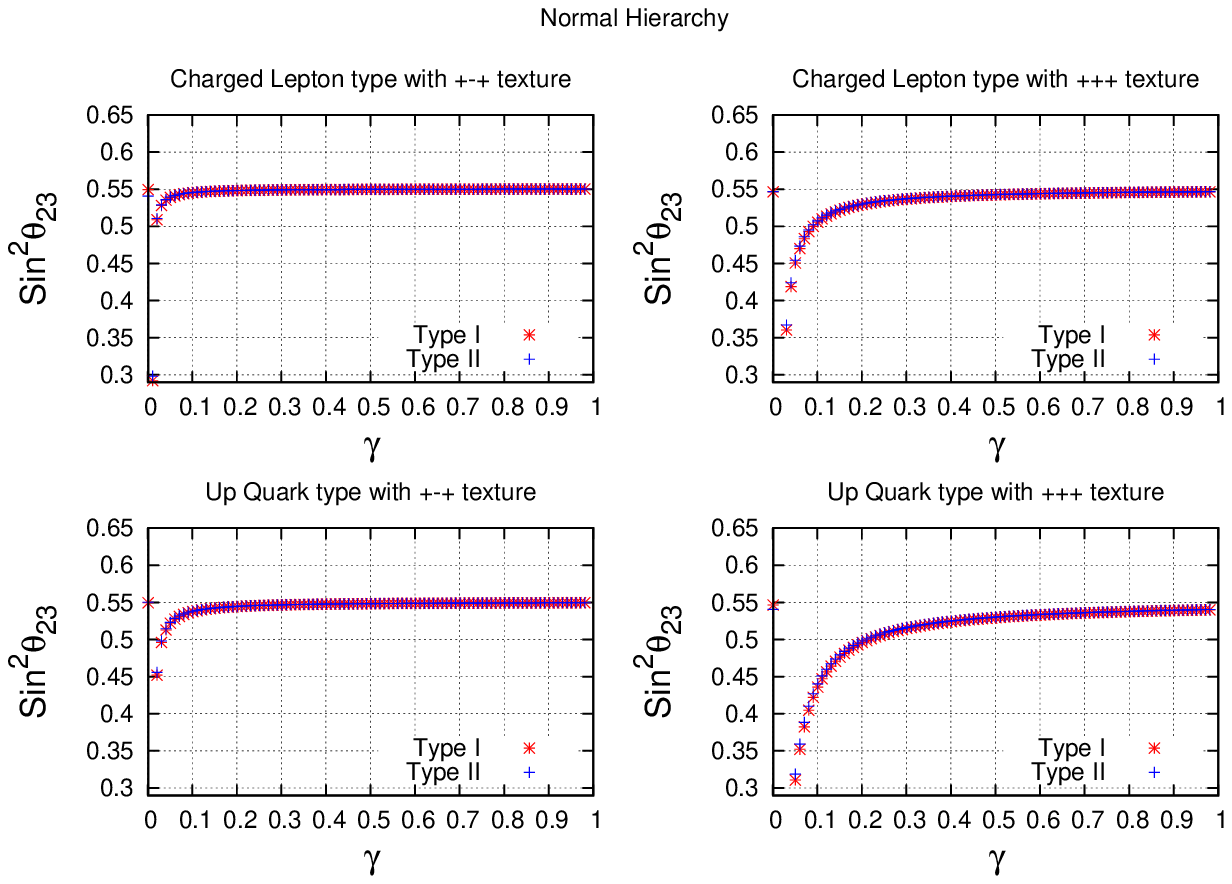}
\caption{Variation of the predicted values of $\sin^2{\theta_{23}}$ as a function of $\gamma$ in NH case}
\label{fig4}
\end{figure}
\begin{figure}[ht]
 \centering
\includegraphics{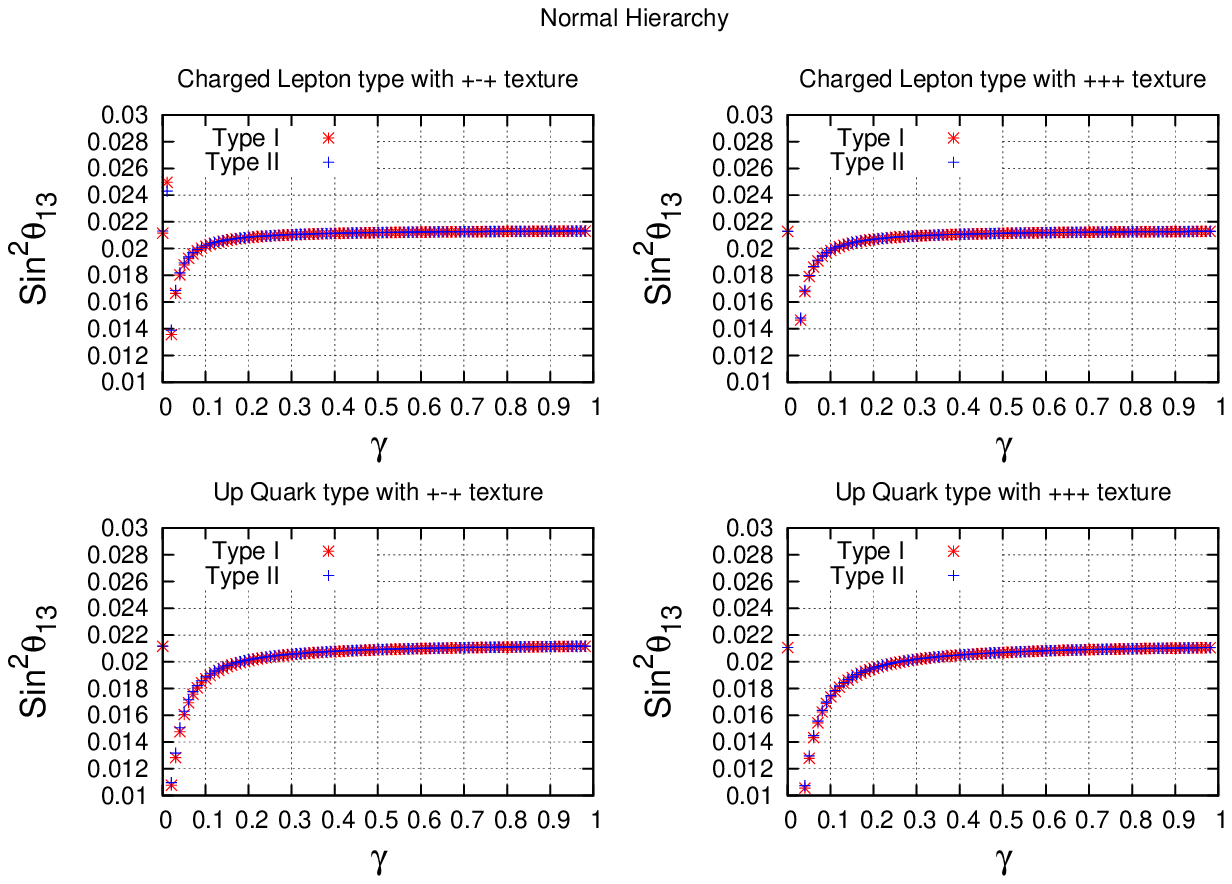}
\caption{Variation of the predicted values of $\sin^2{\theta_{13}}$ as a function of $\gamma$ in NH case}
\label{fig5}
\end{figure}
\begin{figure}[ht]
 \centering
\includegraphics{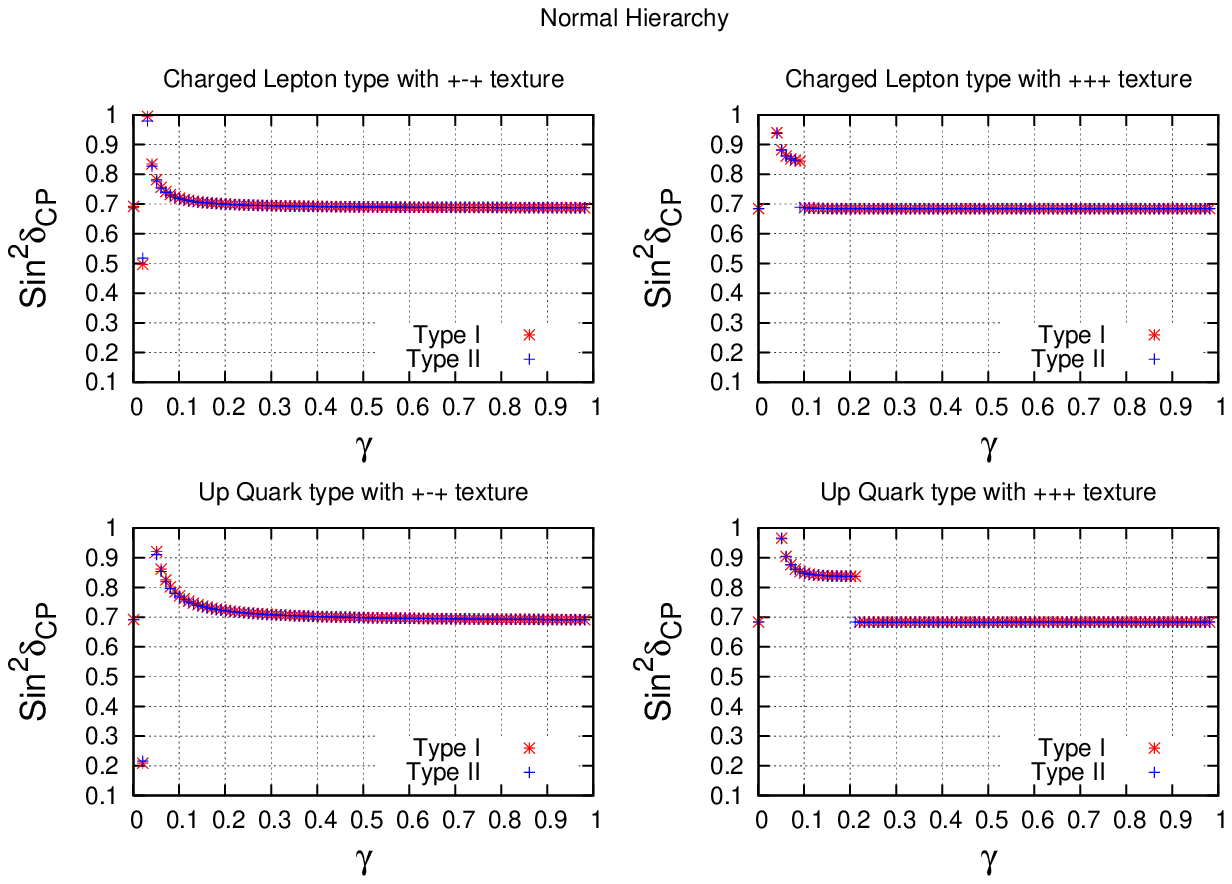}
\caption{Variation of the predicted values of $\sin^2{\delta_{CP}}$ as a function of $\gamma$ in NH case}
\label{fig6}
\end{figure}

For case II i.e., type II dominating seesaw, we calculate $M_{RR}/\gamma$ first using equation (\ref{t2dm}). This fixes the first term in the seesaw formula (\ref{type2b}). Since $M_{RR}$ is not fixed in this case and varry with $\gamma$ as 
$$M_{RR} = \gamma \left ( \frac{1\times 10^{15}}{M_W} \right )^2 m_{LL} $$
the second term in the seesaw formula (\ref{type2b}) also varies with $\gamma$ with minimum contribution for $\gamma \sim 1$. Similar to the case I, here also we vary this dimensionless parameter $\gamma$ from $0.001$ to $1.0$ and check the variations of all the neutrino oscillation parameters.

\begin{figure}[ht]
 \centering
\includegraphics{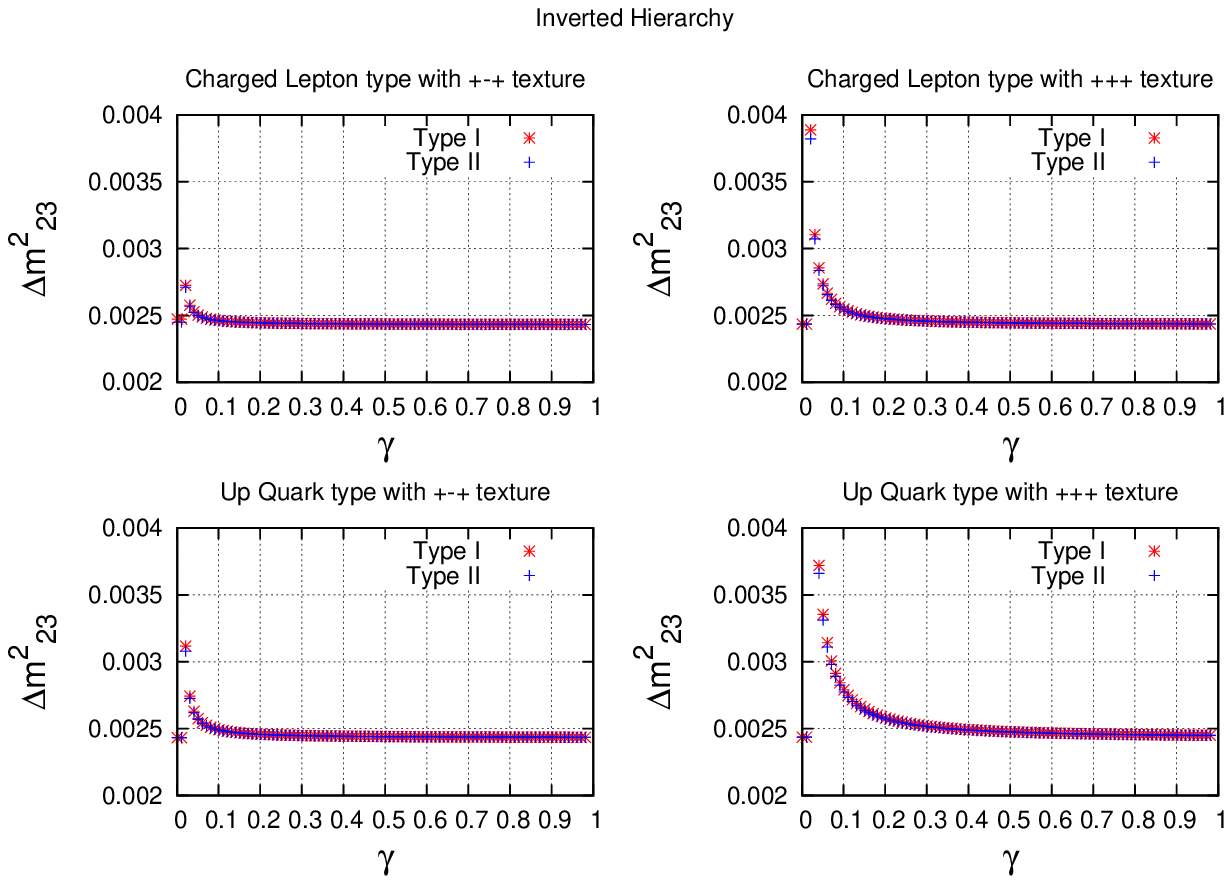}
\caption{Variation of the predicted values of $\Delta m^2_{23}$ as a function of $\gamma$ in IH case}
\label{fig7}
\end{figure}

\begin{figure}[ht]
 \centering
\includegraphics{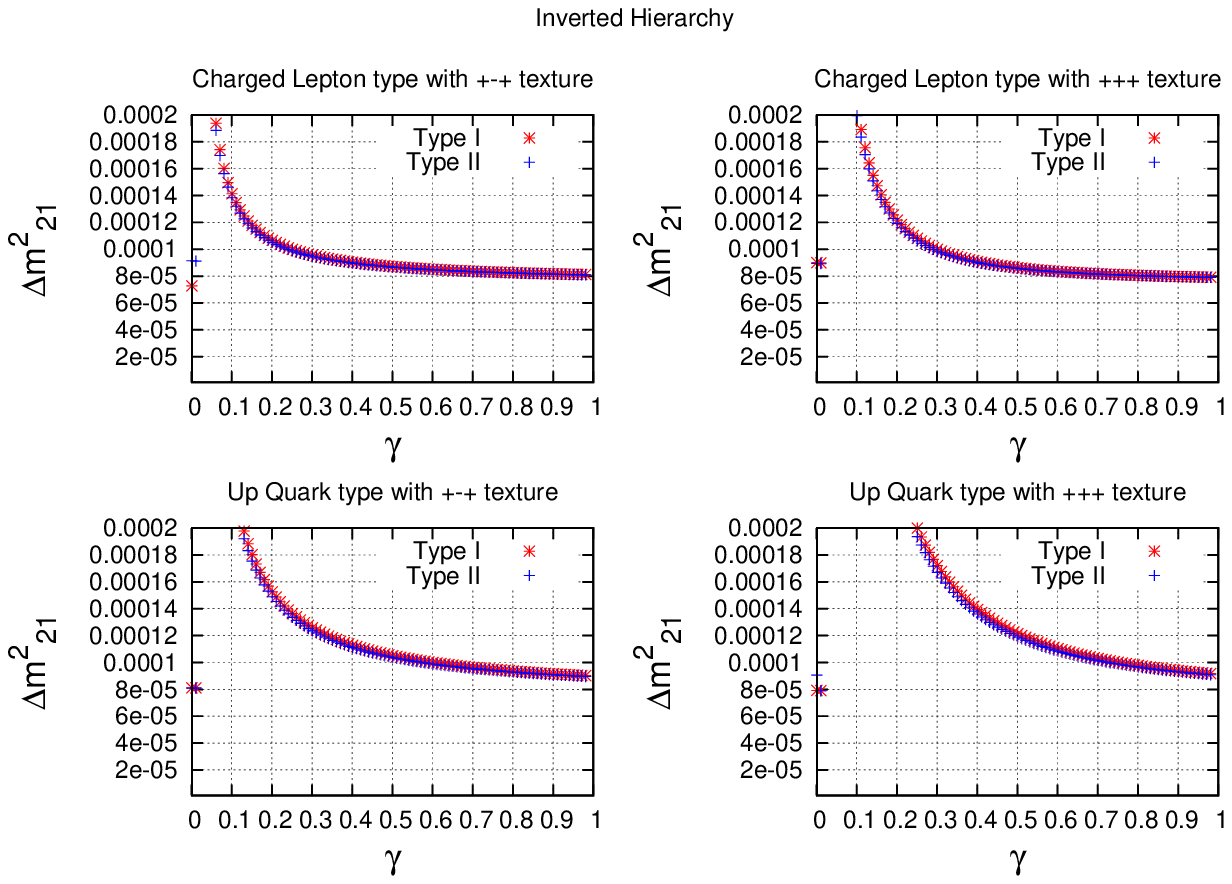}
\caption{Variation of the predicted values of $\Delta m^2_{21}$ as a function of $\gamma$ in IH case}
\label{fig8}
\end{figure}

\begin{figure}[ht]
 \centering
\includegraphics{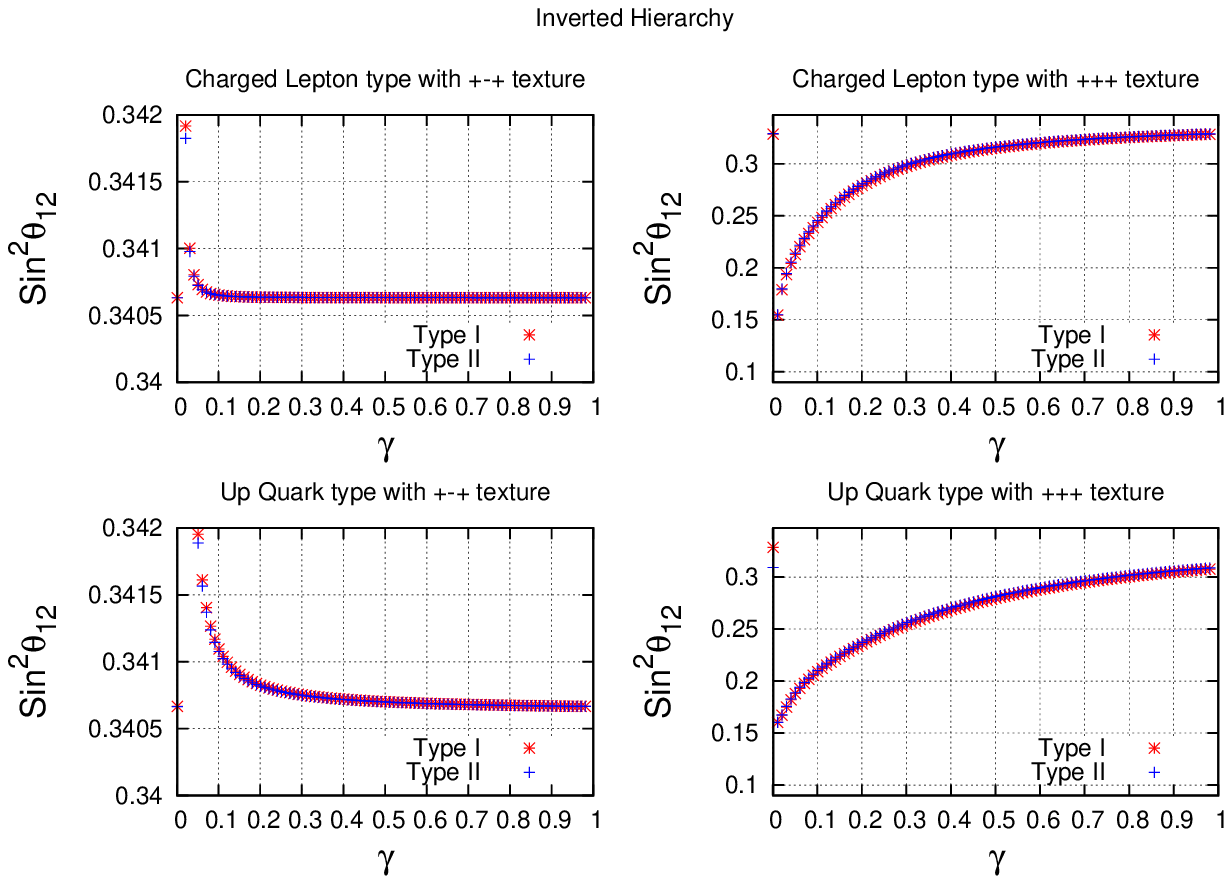}
\caption{Variation of the predicted values of $\sin^2{\theta_{12}}$ as a function of $\gamma$ in IH case}
\label{fig9}
\end{figure}

\begin{figure}[ht]
 \centering
\includegraphics{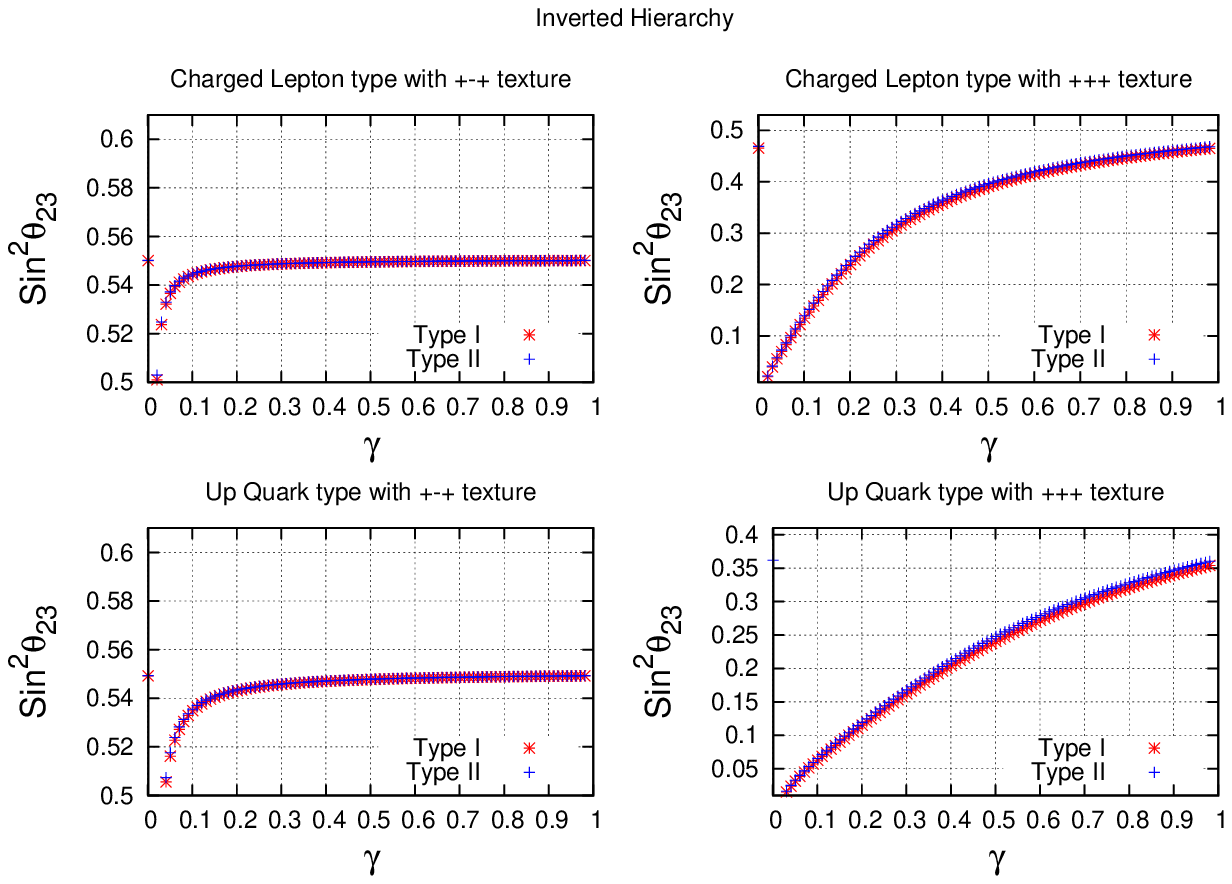}
\caption{Variation of the predicted values of $\sin^2{\theta_{23}}$ as a function of $\gamma$ in IH case}
\label{fig10}
\end{figure}

\begin{figure}[ht]
 \centering
\includegraphics{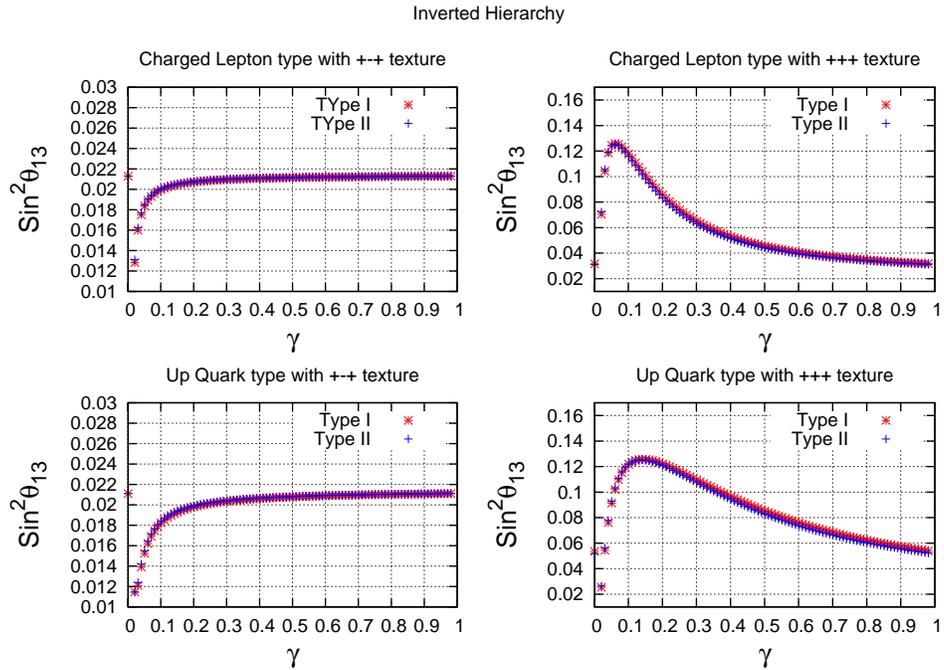}
\caption{Variation of the predicted values of $\sin^2{\theta_{13}}$ as a function of $\gamma$ in IH case}
\label{fig11}
\end{figure}

\begin{figure}[ht]
 \centering
\includegraphics{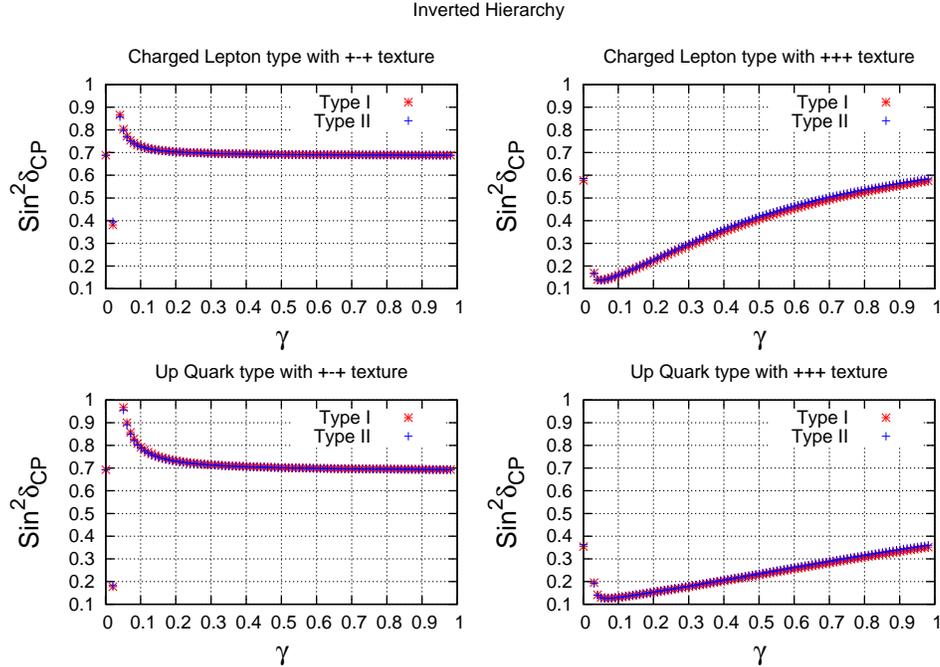}
\caption{Variation of the predicted values of $\sin^2{\delta_{CP}}$ as a function of $\gamma$ in IH case}
\label{fig12}
\end{figure}

The predictions for neutrino oscillation parameters at $\gamma = 1.00$ (which corresponds minimum contribution of the perturbation seesaw term and maximum possibility for the model to survive) for both case I and case II are shown in table \ref{table:results2} and \ref{table:results3}. Here the notation $(+++)$ and $(+-+)$ corresponds to the neutrino mass eigenvalues $(m_1, m_2, m_3)$ and $(m_1, -m_2, m_3)$ respectively. The parameter values which lie outside the $3\sigma$ range of global fit data are shown in red. The results clearly show that the predictions for the parameters do not change substantially if we go from type I dominating case to type II dominating one. The same observation follows from the figures \ref{fig1}, \ref{fig2}, \ref{fig3}, \ref{fig4}, \ref{fig5}, \ref{fig6}, \ref{fig7}, \ref{fig8}, \ref{fig9}, \ref{fig10}, \ref{fig11} and \ref{fig12} where we have shown the variation of the neutrino oscillation parameters with respect to $\gamma \in (0.001, 1.0)$ for all the cases of interest. The variation of the oscillation parameters with respect to $\gamma$ are almost identical for both type I dominating and type II dominating cases with one overlapping the other in most of the cases as seen from the figures.

Apart from the observation that predictions for neutrino oscillation parameters remain almost same for both type I and type II dominating cases, we also observe that certain combinations of IH or NH, Majorana CP phase pattern $(+-+)$ or $(+++)$, $m_{LR}$ of type CL or UQ do not give rise to correct $3\sigma$ predictions of oscillation parameters at $\gamma \sim 1$ where according to our selection criteria, the model is most likely to survive. Such combinations are
\begin{itemize}
\item IH with Majorana CP phase pattern $(+-+)$ for CL type $m_{LR}$ with dominant type I seesaw(Table \ref{table:results2}).
\item IH with Majorana CP phase pattern $(+++)$ for CL type $m_{LR}$ with dominant type I seesaw(Table \ref{table:results2}).
\item IH with Majorana CP phase pattern $(+-+)$ for UQ type $m_{LR}$ (Table \ref{table:results2}).
\item IH with Majorana CP phase pattern $(+++)$ for UQ type $m_{LR}$ (Table \ref{table:results2}).
\item NH with Majorana CP phase pattern $(+-+)$ for UQ type $m_{LR}$ (Table \ref{table:results3}).
\item NH with Majorana CP phase pattern $(+++)$ for UQ type $m_{LR}$ (Table \ref{table:results3}).
\end{itemize}

Thus, the UQ type texture of Dirac neutrino mass matrix is disfavored in our analysis for both types of hierarchies and Majorana CP phase patterns. For CL type Dirac neutrino mass matrix, normal hierarchy is compatible with both types of Majorana CP phase patterns whereas inverted hierarchy gives correct predictions only for the case of type II seesaw dominance.

Also, for these favored models IH(+-+), IH(+++), NH(+-+) and NH(+++) with CL type $m_{LR}$, the predictions for $\Delta m^2_{23} (\text{IH}), \Delta m^2_{31} (\text{NH})$ remain very close to the best fit central values whereas IH prefers $\Delta m^2_{21}$ to be at the upper end of the $3 \sigma$ range. Regarding the mixing angles, $\sin^2{\theta_{23}}$, $\sin^2{\theta_{13}}$ stay very close to the best fit central value for all the models whereas $\sin^2{\theta_{12}}$ stays at the lower end of $3\sigma$ range for NH(+++) model and upper end of $3\sigma$ range for IH(+-+), IH(+++), NH(+-+) models. Therefore, it is very likely that more precise future data from neutrino oscillation experiments will rule out some (if not all) of the scenarios discussed in this work.

\clearpage
\section{Conclusion}
\label{conclude}
We have studied the survivability of neutrino mass models within the framework of generic left-right symmetric models by considering both types of hierarchies (normal and inverted), two types of extremal Majorana CP phases $(+-+)$ and $(+++)$, and two types of Dirac neutrino mass matrices: charged lepton type and up quark type. In generic LRSM, neutrino mass can get contributions from both type I as well as type II seesaw terms. We check that for generic choices of left-right symmetry breaking scale and Dirac neutrino mass matrices, these two different contributions to neutrino mass can be comparable only if a dimensionless parameter appearing in the type II term ($\gamma$ in our notation) is fine tuned to be very small. Without considering this special fine-tuned case, we consider two other possible cases: one in which type I seesaw dominates and the other in which type II seesaw dominates.  
We use the generic parametrization of neutrino mass matrix(TBM plus non leading corrections giving rise to non-zero $\theta_{13}$) obtained by several groups and find the numerical value of these parameters using the global fit neutrino oscillation data as well as the cosmological upper bound on the sum of absolute neutrino masses. We then keep the dominating seesaw term fixed in the seesaw formula and vary the other term by varying the dimensionless parameter $\gamma$. We then calculate the predictions for neutrino oscillation parameters for different values of $\gamma$ and check whether they agree with the $3\sigma$ range of global fit data at $\gamma \sim 1$ which corresponds to the case where the non-leading term in the seesaw formula has the minimum possible contribution. 

Apart from the observation that both type I dominating and type II dominating cases give almost identical predictions, we also observe that up quark type Dirac neutrino mass matrix is disfavored within our framework. Predictions for neutrino parameters using up quark type $m_{LR}$ with both IH, NH and $(+-+)$, $(+++)$ deviate from the $3\sigma$ range of global fit data. In the case of charged lepton type $m_{LR}$, normal hierarchy survives with both types of Majorana CP phase patterns whereas inverted hierarchy survives only with the $(+-+)$ case.
In view of above, the neutrino mass models considered in our study can survive in nature within the framework of type I and type II seesaw mechanism and hence can not be ruled out yet apart from certain exceptions mentioned above. However, some of the model predictions lie at the extreme ends of the $3\sigma$ allowed range and hence more precise data from neutrino oscillation experiments should be able to to rule out some (if not all) of the scenarios we have discussed in our work.

\end{document}